# Ptychography intensity interferometry imaging

Wentao Wang, Qi Han, Hui Chen*, Yuan Yuan, and Zhuo Xu

Intensity interferometry (II) exploits the second-order correlation to acquire the spatial frequency information of an object, which has been used to observe distant stars since 1950s. However, due to unreliability of employed imaging reconstruction algorithms, II can only image simple and sparse objects such as double stars. We here develop a method that overcomes this unreliability problem and enables imaging complex objects by combing II and a ptychography iterative algorithm. Different from previous ptychography iterative-type algorithms that work only for diffractive objects using coherence light sources, our method obtains the objects spatial spectrum from the second-order correlation of intensity fluctuation by using an incoherent source, which therefore largely simplifes the imaging process. Furthermore, by introducing loose supports in the ptychography algorithm, a high-fdelity image can be recovered without knowing the precise size and position of the scanning illumination, which is a strong requirement for traditional ptychography iterative algorithm..

## I. INTRODUCTION

In nowadays, there are many imaging methods which assist people to give a deep vision to the surroundings. Generally, these imaging methods can be sorted into two types in physics. One is the classical imaging that is based on the electromagnetic waves amplitude interferometry (AI), where many waves are superimposed causing the phenomenon of interference to extract information. It can be traced back to the discovery of the wave-property of light carried by Young(1800). And then, Mach-Zehnder and Michelson proposed two interferometers to observe the interference pattern or measure the phase shift, the first-order correlation of the source, by modulating the path difference of two beams splitting from a single source[1]. The other is intensity interferometer imaging, which exploits the high-order correlation knows as photon bunching of the photons based on the intensity interferometry(II) schematic[2-4]. Since the measured information is called correlation function which is the Fourier transform of the average distribution of the object. One can reconstruct its image by taking the inverse Fourier transform of the correlation function. In essence, II is the Hanbury Brown and Twiss(HBT) effect, which utilize the second-order correlation of the intensity fluctuation of two detectors to induce the angular diameter of hot stars with an excellent resolution(microarcsecond)[5-8]. And then, II was used to long baselines imaging to reveal details across and outside stellar surfaces. In classical imaging, it is troubled by the receiving phase accuracy brought by the complex observe environment, such as turbulence. Many methods are adopted to decrease the wavefront distortions, such as adaptive optics, which correct the deformations of an incoming wavefront by deforming a mirror in order to compensate for the distortion[9]. These methods bring the complexity at the same time and each sub-system add its own noise to the whole system and increase the chance for a failure to occur due to the multiplicative effects of errors. However the setup of the II is simple and less dependent on the equipment. And the resolution of this approach is depend on the area of the speckles rather than depending on the size of the imaging lens. The higher resolution requires larger and more perfect lens. This is at a high cost. Furthermore II is immune to the equipment imperfections and the atmospheric turbulence[10-11]. When II is applied to imaging, the main shortcoming is the phase retrieval algorithm. Because the algorithm is tend to trap in local minimum and slow to converge the true image.

Phase retrieval is to reconstruct the phase information from intensity measurements without the reference beam such as holography. The earliest phase retrieval is G-S algorithm proposed by Gerchberg and Saxton[12]. Then it was developed to Error Reduction (ER) and Hybrid Input-Output (HIO) algorithms by Fienup[13-17]. Both two above algorithms are widely accepted and utilized. However, above algorithms are plagued by some fundamental hardships. First their field of views are too much small; secondly, they tend to stuck in the stagnation when facing the objects with complicated structure. To improve the convergence speed and reliability, Rodenburg proposed the ptychography to expand field of view[18]. This method increase the obtained information by overlap illuminating the object. Even the complicated structure can be reconstructed directly[19]. This is because the overlap illumination increase the known object's information and the overlap positions convey the interference which help to converge the unique value fast. However, this method is sensitive to the errors in shape or shift of the probe[20]. And current researches focus on the microscopic which belongs to the classical imaging. And the acquisition equipment is so high precise and complicated that the noise is inevitable[21].

Here we demonstrate II with ptychography to explore the second order correlation imaging in the near field. Firstly, the theory is briefly derived to demonstrate the relationship between the intensity distribution and the correlation function. Secondly, simulation and the experiment is demonstrated. The object is illuminated by the incoherent light and create the speckle pattern. The correlation function is derived from the speckle pattern. Then

* chenhui@mail.xjtu.edu.cn

Electronic Material Research Laboratory, Key Laboratory of the Ministry of Education and International Centre for Dielectric Research, Xi'an Jiaotong University, Xi'an, China, 710049 the object's image is reconstructed by the retrieval algorithm. Thirdly, the phase retrieval algorithm is demonstrated. At last by introducing loose supports in the ptychography algorithm, a high-fidelity image can be recovered without knowing the precise size and position of the scanning illumination, which is a strong requirement for traditional ptychography iterative algorithm.

## II. PRINCIPLE AND SIMULATION

A typical intensity interferometer schematic is composed of two detectors and one correlator. Two detectors placed in one coherent area measure the light intensities from the same source. The correlation between these two detections is an ensemble average of the product of two intensities $I_1$ and $I_2$, which is represented as blow

$$
\begin{aligned}
(I_1(t_1,x_1)I_2(t_2,x_2)) &= (E_1(t_1,x_1)E_1^*(t_1,x_1)\ldots \\
&\quad * E_2^*(t_2,x_2)E_2^*(t_2,x_2)) \\
&= (I_1(t_1,x_1))(I_2(t_2,x_2))(1+|\gamma_{12}|^2)
\end{aligned} \quad (1)
$$

where $E_i(t_i,x_i)$ and $E_i^*(t_i,x_i)$ ($i = 1, 2$) are a pair of conjugate electronic field variables, $I_i(t_i,x_i)$ is the instantaneous intensity at time $t_i$ at detector $d_i$, $\gamma_{12}$ is the mutual correlation function of light intensity between points $x_1$ and $x_2$, and () denotes time ensemble average. According to the van Cittert-Zernike theorem, correlation function $\gamma_{12}$ is the Fourier transform of intensity distribution function $O(\alpha, \beta)$ of the object,

$$
\gamma(\Delta x, \Delta y) = \frac{e^{-j\varphi}\iint O(\alpha,\beta)exp[j\frac{2\pi}{\lambda z}(\Delta x \alpha+\Delta y \beta)]d\alpha d\beta}{\iint I(\alpha,\beta)d\alpha d\beta}. \quad (2)
$$

By Eqs (1) and (2), the correlation function of two detectors can be written as:

$$(I_1(t_1,x_1)I_2(t_2,x_2)) \propto |F\{O(\alpha,\beta)\}|^2. \quad (3)$$

where $O(\alpha, \beta)$ is the intensity distribution function of the object, $(\Delta x, \Delta y)$ is the relative position of the two detectors, $\varphi$ is the phase difference induced by different optical length and $z$ is the distance between object and observer surface. From Eq (3), it is equivalent to perfect plane wave illuminating objects in first-order interference configuration. Therefore, in II, simple manufactures are feasible to obtain the diffraction pattern. However, what measured is the square of the autocorrelation function, losing the phase information, which could not recover the object by means of direct inverse-Fourier transform of Eq (2). Hence, phase retrieval algorithm is necessary to obtain the image of the object. The main approaches that generally applied in phase-retrieval algorithm are Cauchy-Riemann [25] and iterative Fourier phase recovery [26]. In this paper, ptychographical reconstruction algorithm is adopted as one of the iterative Fourier phase recovery [26]. By phase retrieval, the intensity distribution of the object in Eq (2) can be reconstructed.

We firstly give the simulation about the Ptychographical intensity interferometry shown in Fig. 1. The employed source is thermal light, which is classified as incoherent light whose intensity is Rayleighly distributed and phase is uniformly distributed on the interval$(-\pi, \pi)$. And it obeys the circular complex Gaussian statistical distribution. The light illuminates the object and produced the speckle pattern. Then the speckle pattern is recorded to derive the autocorrelation function. At last, by the series autocorrelation functions corresponding to the illumination, the objects image is reconstructed by the phase retrieval algorithm.

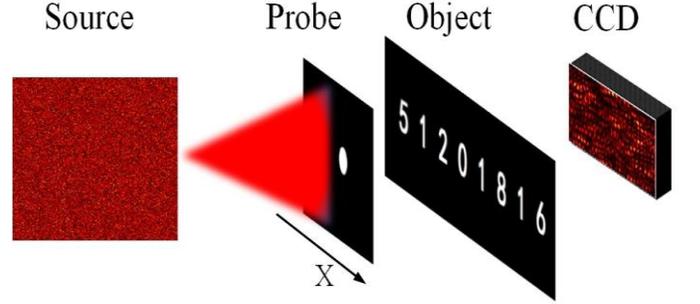

FIG. 1. The arrangement of the simulation. The source is random and obeys the circular complex Gaussian statistical distribution. The probe moves along the X direction to scan the object. The CCD captures the speckle pattern after every illumination.

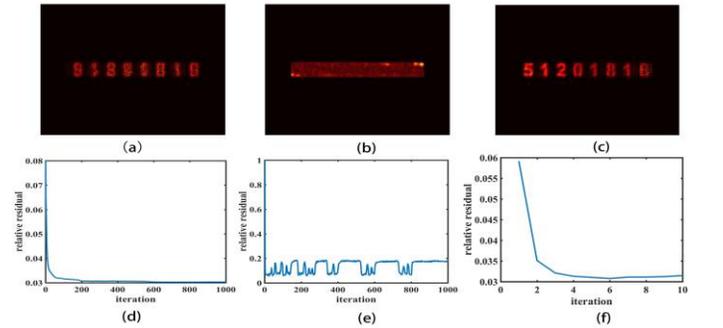

FIG. 2. (a) is the reconstructed image with 1000 ER iterations ; (b) is the reconstructed image with 1000 ER iterations with beta=0.7 ; (c) is the reconstructed image after 10 PII iterations . (d), (e) and (f) are the relative residual at each iteration corresponding to the above retrieval algorithm in reciprocal space, respectively.

In the simualtion, a comparison about three phase retrieval algorithms is shown in Fig. 2. Here speckle pattern are recorded to derive the autocorrelation function. Then the autocorrelation function is sent into ER, HIO and Pty to reconstruct the image, respectively. The result is shown in Fig. 2. (a) is the reconstructed image after 1000 ER iterations. Though the rough sketch is clearly shown, one cannot distinguish the details; (d) is the corresponding relative residual curve at each iteration in reciprocal space. It can be seen that it drops quickly within 100 iterations and then turns out to be steady around 0.03 within following iterations. That means even increasing the iteration number the quality of the reconstruction cannot be improved. The HIO is called to reconstruct the object, the reconstructed image is shown in Fig. 2(b) and the relative residual curve is shown in Fig. 2(e). The reconstructed image is so blurred that nothing can be obtained. The relative residual drops quickly to 0.1 around 50 iterations. At last, the ptychgraphy intensity correlation imaging is called, the reconstructed results are shown in Fig. 2(c) and Fig. 2(f). The image is recovered after 10 iterations and can be clearly distinguished. And the corresponding relative residual curve drops and stays 0.03 after 6 iterations.

## III. EXPERIMENT

The schematic of the experiment is shown in Fig. 5. The laser employed is solid-state and wavelength is 457nm(MG: 85-BLS-601). The laser incident on the reverse-telescope group(Lens1 and Lens2) to be 10 times amplified, then the amplified beam whose diameter is about 20mm incidents on the rotating ground glass (G.G) to produce the pseudothermal light. The voltage on the ground glass is 12v. The speed of the ground glass is 0.6 rpm(round per miniute). Then the pseudothermal light propagates to the probe to illuminating the object. The distance between the G.G and the probe is 300mm. So the coherence length on the probe plane is about $6.855\mu m$ according to $\Delta x = \lambda * z/D$, where $\lambda$ is the center wavelength of incident light, z is the propagation distance and D is the typical size. Finally the speckle pattern is recorded by the CCD which locates at 250mm after the object. The diameter of the probe applied in the experiment is 5mm and move along the X direction with each step 0.5mm. The number of the steps is 16. And CCD captures 1000 different speckle patterns at each step and the exposure time is 20ms. The coherence length of the speckle is about $23\mu m$ according to the above equation. The computer displays one captured speckle pattern at step 1. The number of speckle pattern here used to calculate the autocorrelation function is 1000.

Then the probe is set be front of the CCD to explore the long-distance observation just shown in Fig. 5(b). The results is shown in Fig. 8. The reconstructed object is some inclined, this is because of the object settlement in the experiment. Here, we are successful to demonstrate ptychographical intensity interferometry where the probe is located in the front of the object and detector respectively by employing the pseudothermal light. However, the precise size of the probe movement is obtained from the translation stage in the lab and there is some bias about the probe movement in real application. Hence we give some discussion about the fault tolerance of the movement in our retrieval algorithm.

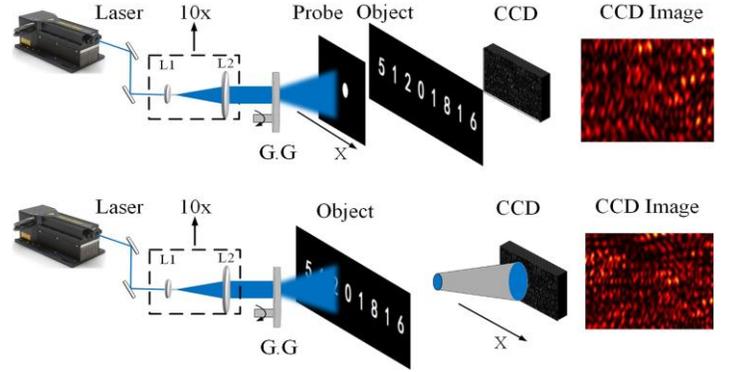

FIG. 3. The schematic of the experiment setup. (a) the probe is in the front of the object, this setup is used for short distance, the wavelength of the laser is 457nm; the focal length of lens1 is 25.4mm; the focal length of lens2 is 250mm; G.G is the rotating ground glass; probe is the circular hole with diameter 5mm; the computer displays the captured speckle pattern by CCD. (b) the probe is behind of the object, this setup is used for long distance.

## IV. RECONSTRUCTION METHOD

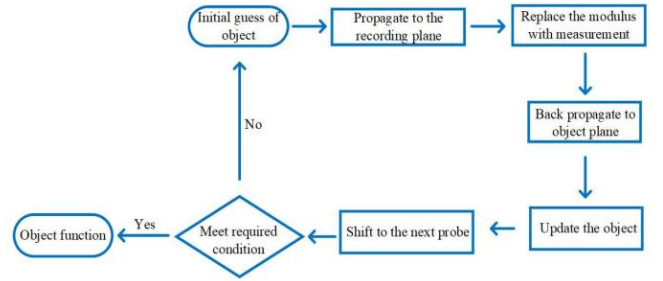

FIG. 4. Flowchart of the reconstruction process with the method in the paper.

The data flowchart of the reconstruction is schematically shown in Fig. 1, where the iterative reconstruction is carried out after the initial guess is given to the object function:

(1) The object $O(r)$ is illuminated by the light souce $P(r - R)$ which is modulated by the probe $P(r)$ to produce the light field $X(r, R) = O(r)P(r - R)$. The object is assumed to be very thin.

(2) The light field $F(k)$ on the CCD plane is numeri-

cally propagating the *X*(*r*, *R*) to the recording plane and replacing its modulus with the square root of the corresponding calculating autocorrelation function *amp*(*r*−*R*) and the phase information keep unchanged.

(3) The modified *F*(*k*) is then propagated to the object plane to obtain the updated the object plane $O_n(r)$ with the following equation:

$O_n(r) = O_{n-1}(r) + X_n(r, R)$.

(4) The above calculation is sent to the next illumination probe until the above calculation has made at all the illumination positions.

(5) The iteration will stop after achieve the maximum iteration number.

## V. DISSCUSSION

Here we analyse the effect of shift error for the ptychographical intensity interferometry. The experiment setup is the same as Fig. 6(a). The reconstructed images are shown in Fig. 8. When there is no shift error, we can successfully obtain the object image. The reconstructed Then increasing the shift to 20%, the reconstructed image is becoming little blurred. Under 25% shift error, it is hard to distinguish the object information. At last it is becoming so mess that no information is distinguished from the reconstructed image under 50% shift error.

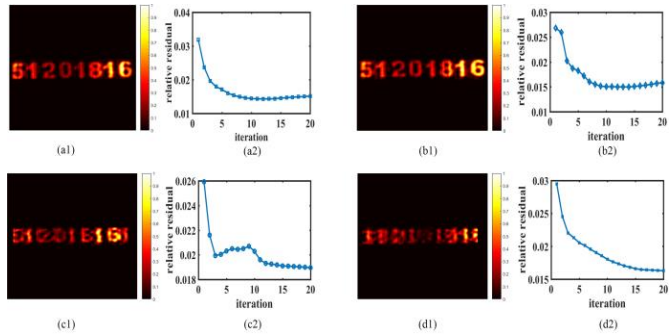

FIG. 5. The reconstructed image with 20 iterations under different shift errors: (a1) reconstruction under no shift error; (a2) the relative residual curve under no shift error; (b1) reconstruction under 10% shift error; (b2) relative residual curve corresponding to 10% shift error; (c1) reconstruction under 25% shift error; (c2) relative residual curve corresponding to 25% shift error; (d1) reconstruction under 50% shift error; (d2) relative residual curve corresponding to 50% shift error.

However it is hard to control the shift error with long distance observation in reality. Here we provide a solution to solve this trouble. In reconstructing process, some looses are given to enlarge the support under 25% and 50% shift error respectively. The result is shown in Fig. 9. For 25% shift error, when we enlarge loose to 5, the reconstructed image is becoming evident. It is clearly distinguish under 10 and 15 enlarge respectively. To enlarge loose to 20, the reconstructed image becomes so mess to distinguish the details. For 50% shift error, when enlarge loose to 15 and 20 respectively, the reconstructed images are clearly distinguish the details. Here we can conclude that when there is shift error, it is possible to adjust the loose of the support to reconstruct the evident image.

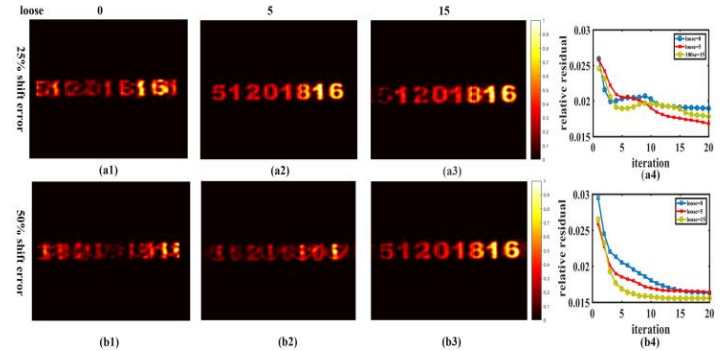

FIG. 6. The reconstructed images under different support looses with 20 iterations. The first row is corresponding to the 25% shift error and the second row corresponding to the 50% shift error. (a1), (a2), (a3) are reconstructed images corresponding to support loose = 0, 5, 15 respectively under 25% shift error; (a4) the relative residual curve under above three situations; (b1), (b2), (b3) are reconstructed images corresponding to support loose = 0, 5, 15 respectively under 50% shift error; (b4) the relative residual curve under above three situations.

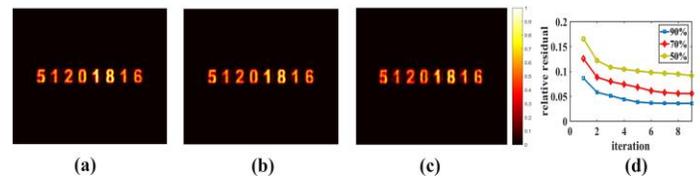

FIG. 7. The reconstructed images under different support looses with 20 iterations. The first row is corresponding to the 25% shift error and the second row corresponding to the 50% shift error. (a1), (a2), (a3) are reconstructed images corresponding to support loose = 0, 5, 15 respectively under 25% shift error; (a4) the relative residual curve under above three situations; (b1), (b2), (b3) are reconstructed images corresponding to support loose = 0, 5, 15 respectively under 50% shift error; (b4) the relative residual curve under above three situations.

## VI. CONCLUSION

In summary, we have proposed and demonstrated an imaging method, termed ptychographical intensity interferometry. This approach reconstruct a sample image from many speckle patterns with the incoherent illumination. It is the first time that proposes and realises the ptychographical intensity correlation imaging in the lab. The use of the ptychography intensity interferome-try framework for reconstructing a sample image is new and may provide an alternative solution for the existing Cherenkov Telescope Array to observe the long distance stars.

Several advantages associated with the reported method is stated as below. Firstly, it increases the reliability when reconstructing the complicated object that the traditional phase retrieval algorithm cannot cooperate. Secondly, it improve the error toleration of the probe shift by adjusting the loose of the support. At last the this method is very efficient in terms of compution cost. The solution typically converges with 5-20 loops. In our experiment, we used 11-21 loops to reconstruct the images. There are several limitations associated with the reported ptychography intensity interferometry. Firstly the autocorrelation is derived from large number of speckle patterns. This increase the collecting time and computational time. In the experiment, we found that autocorrelation derived from 10 speckle patterns can reconstruct the sample at least shown in Fig. 7. Secondly the pixel size of CCD employed in the experiment is 6.45$\mu m$, this restricts the measured speckle size and the resolution of the sample is restricted in hence. Thirdly the minimum exposure time of the CCD is about 0.1$ms$, whcih is much longer than the coherent time of thermal source($fs$). Hence it cannot capture the intensity fluctuations of thermal source.

## REFERENCES

[1] M. Born and E. Wolf. Principles of optics: electromagnetic theory of propagation, interference and diraction of light. Cambridge Univ. Press, Cambridge, 7. edition, 2006.
[2] R. H. Brown, R. Q. Twiss, Nature. 177, 27–29 (1956).
[3] R. H. Brown, R. Q. Twiss, Proc. R. Soc. A. 248, 222–237 (1958).
[4] R. H. Brown, The intensity interferometer (Taylor&Francis LTD, 1974).
[5] E. Brannen, H. I. S. Feguson, Nature. 178, 4548 (1956).
[6] R. H. Brown, R. Q. Twiss, Proceedings of the Royal Society A. 242, 1230 (1957).
[7] R. Hanbury Brown, R. Q. Twiss, Proceedings of the Royal Society A. 243, 1234 (1958).
[8] R. J. Glauber, Phys. Rev, 130, 2529-2539 (1963).
[9] Beckers, J.M, "Adaptive Optics for Astronomy: Principles, Performance, and Applications". Annual Review of Astronomy and Astrophysics. 31 (1): 1362(1993).
[10] P. Nunez, Towards Optical Intensity Interferometry for High Angular Resolution Stellar Astrophysics, University of Utah (2012)
[11] H. Chen, T. Peng, and Y. H. Shih, Phys. Rev. A 88, 023808(2013).
[12] R. W. Gerchberg and W. O. Saxton. A practical algorithm for determination of phase from image and diraction plane pictures. Optik,35(2):237-246, 1972.
[13] J. R. Fienup. Phase retrieval algorithms - a comparison. Appl. Opt., 21(15):2758-2769, 1982.
[14] J. R. Fienup. Reconstruction of an object from the modulus of its Fourier transform. Opt. Lett., 3(1):2729, 1978.
[15] J. R. Fienup. Invariant error metrics for image reconstruction. Appl.Opt., 36(32):8352-2357, Nov. 1997.
[16] J. R. Fienup. Lensless coherent imaging by phase retrieval with an illumination pattern constraint. Opt. Express, 14:498, 2006.
[17] W. Hoppe. Trace Structure-Analysis, Ptychography, Phase Tomography. Ultramicroscopy, 10(3):187-198, 1982.
[18] H. M. L. Faulkner and J. M. Rodenburg. Movable aperture lensless transmission microscopy: a novel phase retrieval algorithm. Phys. Rev. Lett., 93:023903, 2004.
[19] J. M. Rodenburg, A. C. Hurst, A. G. Cullis, B. R. Dobson, F. Pfeier, O. Bunk, C. David, K. Jemovs, and I. Johnson. Hard X-ray lensless imaging of extended objects. Phys. Rev. Lett., 98:034801, 2007.
[20] P. Thibault. Algorithmic methods in diraction microscopy. PhD thesis,Cornell University, August 2007.
[21] J. M. Rodenburg, A. C. Hurst, A. G. Cullis, B. R. Dobson, F. Pfeier, O. Bunk, C. David, K. Jemovs, and I. Johnson. Hard X-ray lensless imaging of extended objects. Phys. Rev. Lett., 98:034801, 2007.